\documentclass[12pt,preprint]{aastex}

\shorttitle{Swift Observations of RS Ophiuchi}
\shortauthors{Bode et al.}

\begin{document}


\title{Swift observations of the 2006 outburst of the recurrent nova RS Ophiuchi: \\ I. Early X-ray emission from the shocked ejecta and red giant wind}


\author{M.F. Bode}
\affil{Astrophysics Research Institute, Liverpool John Moores University, Birkenhead, CH41 1LD, UK}
\email{mfb@astro.livjm.ac.uk}

\author{T.J. O'Brien}
\affil{Jodrell Bank Observatory, School of Physics and Astronomy, University of Manchester, Macclesfield, SK11 9DL, UK}
\email{tob@jb.man.ac.uk}


\author{J.P. Osborne, K.L. Page} 
\affil{Department of Physics and Astronomy, University of Leicester, LE1 7RH, UK}
\email{julo@star.le.ac.uk}
\email{kpa@star.le.ac.uk}

\author{F. Senziani}
\affil{INAF - IASF  Via E. Bassini 15, 20133, Milan, Italy and Universit\'{e} Paul Sabatier, 31062 Toulouse, France}
\email{senziani@lambrate.inaf.it}
 
\author{G.K. Skinner}
\affil{CESR, 31028 Toulouse, France and Universit\'{e} Paul Sabatier, 31062 Toulouse, France}
\email{skinner@cesr.fr}


\author{S. Starrfield, J-U. Ness}
\affil{Dept of Physics and Astronomy, Arizona State University, P.O. Box 871504,Tempe, AZ 85287-1504, USA}
\email{sumner.starrfield@asu.edu}
\email{Jan-Uwe.Ness@asu.edu}
 
\author{J.J. Drake}
\affil{Smithsonian Astrophysical Observatory, 60 Garden Street, MS 3, Cambridge, MA 02138, USA}
\email{jdrake@cfa.harvard.edu} 

\author{G. Schwarz} 
\affil{Department of Geology and Astronomy, West Chester University, West Chester, PA 19383, USA}
\email{gschwarz@as.arizona.edu}

\author{A.P. Beardmore} 
\affil{Department of Physics and Astronomy, University of Leicester, LE1 7RH, UK}
\email{apb@star.le.ac.uk}

\author{M.J. Darnley}
\affil{Astrophysics Research Institute, Liverpool John Moores University, Birkenhead, CH41 1LD, UK}
\email{mjd@astro.livjm.ac.uk}

\author{S.P.S. Eyres} 
\affil{Centre for Astrophysics, University of Central Lancashire, Preston, PR1 2HE, UK}
\email{spseyres@uclan.ac.uk} 

\author{A. Evans} 
\affil{Astronomy Group, School of Physical and Geographical Sciences, Keele University, ST5 5BG, UK}
\email{ae@astro.keele.ac.uk} 

\author{N. Gehrels}
\affil{NASA Goddard Space Flight Center, Greenbelt, MD 20771, USA}
\email{gehrels@milkyway.gsfc.nasa.gov}
 
\author{M.R. Goad} 
\affil{Department of Physics and Astronomy, University of Leicester, LE1 7RH, UK}
\email{mrg@star.le.ac.uk}

\author{P.Jean}
\affil{CESR, 31028 Toulouse, France and Universit\'{e} Paul Sabatier, 31062 Toulouse, France}
\email{jean@cesr.fr} 

\author{J. Krautter}
\affil{Landessternwarte, K\H{o}nigstuhl, 69117 Heidelberg, Germany}
\email{jkrautte@lsw.uni-heidelberg.de} 

\author{G. Novara} 
\affil{INAF - IASF  Via E. Bassini 15, 20133, Milan, Italy}
\email{novara@iasf-milano.inaf.it}




\begin{abstract}

RS Ophiuchi began its latest outburst on 2006 February 12. Previous outbursts have indicated that high velocity ejecta interact with a pre-existing red giant wind, setting up shock systems analogous to those seen in Supernova Remnants. However, in the previous outburst in 1985, X-ray observations did not commence until $55$ days after the initial explosion. Here we report on {\it Swift} observations covering the first month of the 2006 outburst with the Burst Alert (BAT) and X-ray Telescope (XRT) instruments. RS Oph was clearly detected in the BAT 14-25 keV band from $t = 0$ to $t \sim 6$ days. XRT observationsfrom 0.3-10 keV, started at 3.17 days after outburst. The rapidly evolving XRT spectra clearly show the presence of both line and continuum emission which can be fitted by thermal emission from hot gas whose characteristic temperature, overlying absorbing column, $[N_H]_W$, and resulting unabsorbed total flux decline monotonically after the first few days. Derived shock velocities are in good agreement with those found from observations at other wavelengths. Similarly, $[N_H]_W$ is in accord with that expected from the red giant wind ahead of the forward shock. We confirm the basic models of the 1985 outburst and conclude that standard Phase I remnant evolution terminated by $t \sim 10$ days and the remnant then rapidly evolved to display behaviour characteristic of Phase III. Around $t = 26$ days however, a new, luminous and highly variable soft X-ray source began to appear whose origin will be explored in a subsequent paper.

\end{abstract}


\keywords{binaries: close - novae, cataclysmic variables - stars: individual (RS Oph) stars: supernovae - stars: symbiotic - white dwarfs}



\section{Introduction}

\object{RS Ophiuchi} is a Symbiotic Recurrent Nova (RN) which had previously undergone recorded outbursts in 1898, 1933, 1958, 1967 and 1985 \citep[see][]{ros87,rosi87}, with a possible additional outburst in 1907 \citep{sch04}. On 2006 February 12.83UT it was observed to be undergoing a further eruption \citep{hir06}, reaching magnitude V=4.5 at this time. For the purposes of this paper, we define this as $t = t_0$. The optical light curve then continued a rapid decline, consistent with that seen in previous outbursts (\citet{ros87}, AAVSO\footnote{http://www.aavso.org}).

The RS Oph binary system comprises a red giant star in a $455.72\pm0.83$ day orbit with a white dwarf (WD) of mass near the Chandrasekhar limit \citep[see][]{dob94,sho96,fek00}. Accretion of hydrogen-rich material from the red giant onto the WD surface leads to the conditions for a thermonuclear runaway (TNR) in a similar fashion to that for Classical Novae (CNe). The much shorter inter-outburst period for this type of RN compared to CNe is thought to be due to a combination of the high WD mass and a supposed high accretion rate \citep[e.g.][]{sta85,yar05}. Such models lead to the ejection of somewhat lower masses at higher velocities than those for CN models (typically 5000 km s$^{-1}$ and $10^{-8} - 10^{-6}$M$_\odot$ respectively for RNe). Spectroscopy of RS Oph has indeed shown H$\alpha$ line emission with FWHM $= 3930$ km s$^{-1}$ and FWZI $= 7540$ km s$^{-1}$ on 2006 February 14.2 ($t = 1.37$ days, \citet{bui06}). Superimposed on the broad line is an intense and narrow double-peaked structure.

Unlike CNe, where the mass donor is a low-mass main-sequence star, the presence of the red giant in the RS Oph system means that the high velocity ejecta run into a dense circumstellar medium in the form of the red giant wind, setting up a shock system with gas temperatures $\sim 2.2 \times 10^{8}$K for $v_{S} = 4000$ km s$^{-1}$, where $v_S$ is the velocity of the forward shock running into the pre-existing wind (see below). Evidence for the presence of such high temperature material in outbursts prior to 1985 came from observations of coronal lines in optical spectra \citep{ros87} and the expected deceleration was evidenced by the narrowing of the initially broad emission lines \citep{sni87,sho96}. The superimposed narrow lines are then from emission and absorption in the red giant wind ahead of the forward shock. 

The interstellar absorbing column, $N_H = (2.4 \pm 0.6) \times 10^{21}$ cm$^{-2}$, was determined from HI 21cm measurements \citep{hje86} and is consistent with the visual extinction ($E(B-V) = 0.7\pm0.1$) determined from IUE observations in 1985 \citep{sni87}. These quantities, together with the observed versus theoretical bolometric luminosity and apparent brightness of the red giant, are all consistent with a distance to RS Oph of $1.6\pm0.3$ kpc \citep{bods87}.

EXOSAT observations in 1985 \citep{mas87} from $t = 55 - 251$ days post-outburst showed that RS Oph was initially an intense, then a rapidly declining soft X-ray source. \citet{bod85} formulated an analytical model of the evolution of the outburst based on the X-ray emission observed by EXOSAT at 55 days, the rapidly increasing radio emission from $t = 18$ days \citep{pad85} and parameters of the red giant wind derived from optical observations. They concluded that RS Oph evolved like a Supernova Remnant (SNR), but on timescales around $10^5$ times faster (see below). Subsequently, \citet{obr87} and \citet{obr92} constructed detailed analytical and numerical models of the interaction of the ejecta with the circumstellar medium which led to consistent estimates of the outburst energy and ejected mass of $1.1 \times 10^{43}$ erg and $1.1 \times 10^{-6}$ M$_\odot$ respectively, with the ratio of mass loss rate to outflow speed in the red giant wind being estimated as $6 \times 10^{12}$ g cm$^{-1}$, compatible with that from isolated red giants. Perhaps not surprisingly, this
relatively simple model failed to agree fully with the observed spectral
evolution of the X-ray emission. In particular, it was difficult to reconcile
both the low-energy and higher energy behaviour of the emission (see also \citet{ito90} who primarily explored the interaction of shells from two separate outbursts).

\section{{\it Swift} Observations}

{\it Swift} \citep{geh04} is a multiwavelength mission primarily designed to detect gamma-ray bursts (GRBs). However the wide-field, hard X-ray Burst Alert Telescope (BAT, used for the detection of  GRBs, \citet{bar05}), and the X-ray telescope (XRT, used for their follow-up, \citet{bur05}), are proving to be invaluable tools for sky monitoring and for observations of variable astrophysical objects such as novae.

\subsection{BAT Observations}

While awaiting GRBs, BAT normally operates in `survey mode' in which spectral and 
imaging information is available with a time resolution which is typically 300s.  A given part 
of the sky usually falls within the 1.4 sr BAT field of view many times per day. For example, 
64 observations of RS Oph, totalling more than 9 hours, are available in the period 3 days 
either side of the optical detection of the outburst. 

We have analysed all such data from $t = -20$ to +20 days using version 2.3 of the 
{\it Swift} software\footnote{http://swift.gsfc.nasa.gov/docs/software/lheasoft}. Standard filtering was used to remove data affected by bad quality, high 
background rates, source occultations, etc.  Two non-standard programs were used for 
the overlaying of images and for the extraction of source count rates while allowing for 
the presence of  strong sources in the field of view. The latter allows for the intensity of all 
sources in the field of view to be fitted simultaneously, along with detector background 
models. It was used to calculate weighted means of the estimated intensity at the position of 
RS Oph in 4 broad bands (14-25, 25-50, 50-100 and 100-190 keV).

\begin{figure}
\centering
\includegraphics[width=0.8\textwidth,angle=-90]{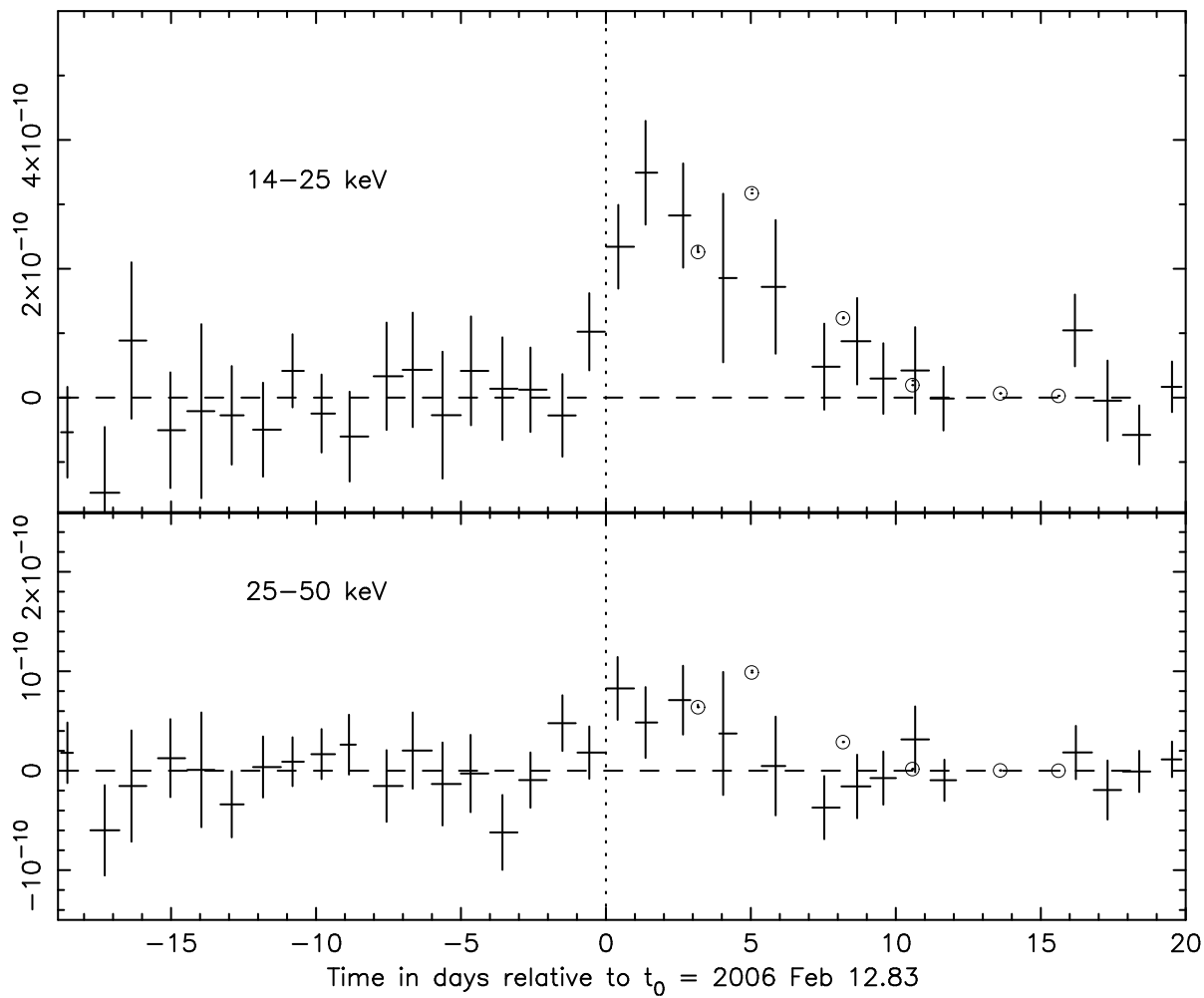}
\caption{The flux from RS Oph in the bands 14-25 keV and 25-50 keV observed with {\it Swift} BAT. Units are ergs 
cm$^{-2}$ s$^{-1}$ and are based on count rates in these bands assuming the 
spectral form found from the observation 001 XRT data. The circles show the flux in the same 
energy bands from  extrapolation of the models in Table \ref{tab1}. Means have been calculated  over 24 h periods and are represented by a 
point at the weighted
mean time, with horizontal error bars stretching from the first to the last sample included in
the mean . The very irregular sampling  leads to uneven horizontal error 
bars. The dotted line is the time of the 
first optical detection of the outburst.  \label{bat}}
\end{figure}

A clear detection of the source was made in the 14-25 keV band during the first 3 days from $t_0$ (Figure \ref{bat}). In an overlay of  images formed from the corresponding data, 
standard software (batcelldetect) finds a  $9.9\sigma$ source within 2.5 arc minutes of the 
position of  RS Oph as the only unidentified source above $5\sigma$, confirming the reality of 
this detection.  There are tantalising hints of emission in this same band  in the 
preceding few days, and also a weak detection in the 25-50 keV band immediately following the 
outburst, but these should be treated with caution and their evaluation awaits a 
more detailed analysis.

Standard software for spectral analysis was used. Spectra with 2 keV bins were extracted for each interval over which the 
pointing was unchanged  (averaging 840 s). Systematic errors\footnote{Per version 20051103 of the 
BAT CALDB}  were taken into account and XSPEC\footnote{http://heasarc/gsfc.nasa.gov/docs/xanadu/xspec} Version 11.3.2  was used to fit models 
simultaneously to all the observations covering $\sim$1 day. The signal to noise ratio was never 
good enough to justify models with multiple free parameters and so only the normalisation 
was fitted. 

Unfortunately, due to a BAT reboot,  no data are available at the time of the first XRT 
observation discussed below, and when data again became available the source was comparatively weak. We have fitted the BAT 14-50 keV spectra with the 
simple  spectral models found using the XRT data as detailed below. Satisfactory $\chi^2$ values 
were always found and allowing for the fact that BAT data are averages, and centered on 
slightly different times, the measured fluxes are consistent with extrapolations of the 
corresponding XRT spectra. 

The RXTE All-sky Monitor \footnote{http://xte.mit.edu} observed the direction of RS Oph at 
several times during the period 2006 February 12-18 ($t \sim$0 to 5 days). Little or nothing is detected in band $a$ 
(1.3-3 keV) while in band $b$  (3-5 keV), and more particularly in band $c$ (5-12 keV),  an 
increasing flux is seen. The fluxes are initially lower than expected from a model based on the first XRT observation, scaled to match the BAT data, but they 
reach the expected level by about February 15. This is consistent with an initially very high, but 
decreasing, absorption column. \citet{sok06} have published an analysis of  RXTE PCA
observations starting from $t = 3.1$ days  The hard X-ray light curve implied by
their results is consistent with Figure \ref{bat}, although the BAT observations extend
higher in energy and the XRT data described below provide better low energy
coverage and energy resolution below 10 keV. 

\subsection{XRT Observations}

Table \ref{tab1} gives details of the {\it Swift} XRT observations carried out over the first 26 days following the outburst. These began at $t=3.17$ days \citep{bod06,bod06a} and continued at a frequency which has given excellent temporal coverage. This 
means that the observations began much earlier in the evolution of the outburst, and we were able to follow it in much more temporal and spectral detail, than was the case for the EXOSAT observations in 1985.

{\it Swift} software version 2.3 was used to process the data from the XRT.  
Source spectra were extracted from the cleaned Windowed Timing mode
event-lists, using a box region of 60x20 pixels (1 pixel = 2.36 arcsec);  
the same region was then offset from the source to obtain a background
spectrum. During some of the observations the source was positioned over
the bad CCD columns \citep[caused by a micrometeorite impact on 2005-05-27]{abb06} and we corrected for the fractional loss of the PSF incurred when
calculating the count rate and flux.

Grade 0-2 events were chosen for the spectral analysis. The {\sc ftool}
{\bf xrtmkarf} was used to generate suitable ancillary response function
files and these were used in conjunction with the most up-to-date response
matrix (swxwt0to2$\_$20010101v007.rmf).

All spectra were grouped to a minimum of 20 counts per bin in order to
facilitate the use of the $\chi^2$ statistic in XSPEC, the results of which are discussed below. All fits included an 
energy scale offset that was allowed to vary between -0.1 and 0 keV to 
account for possible CCD bias measurement uncertanties. Errors are given
at the 90~per~cent level (e.g., $\Delta\chi^{2}$~=~2.7 for one degree of
freedom).

\begin{figure}
\centering
\includegraphics[width=0.75\textwidth,angle=-90]{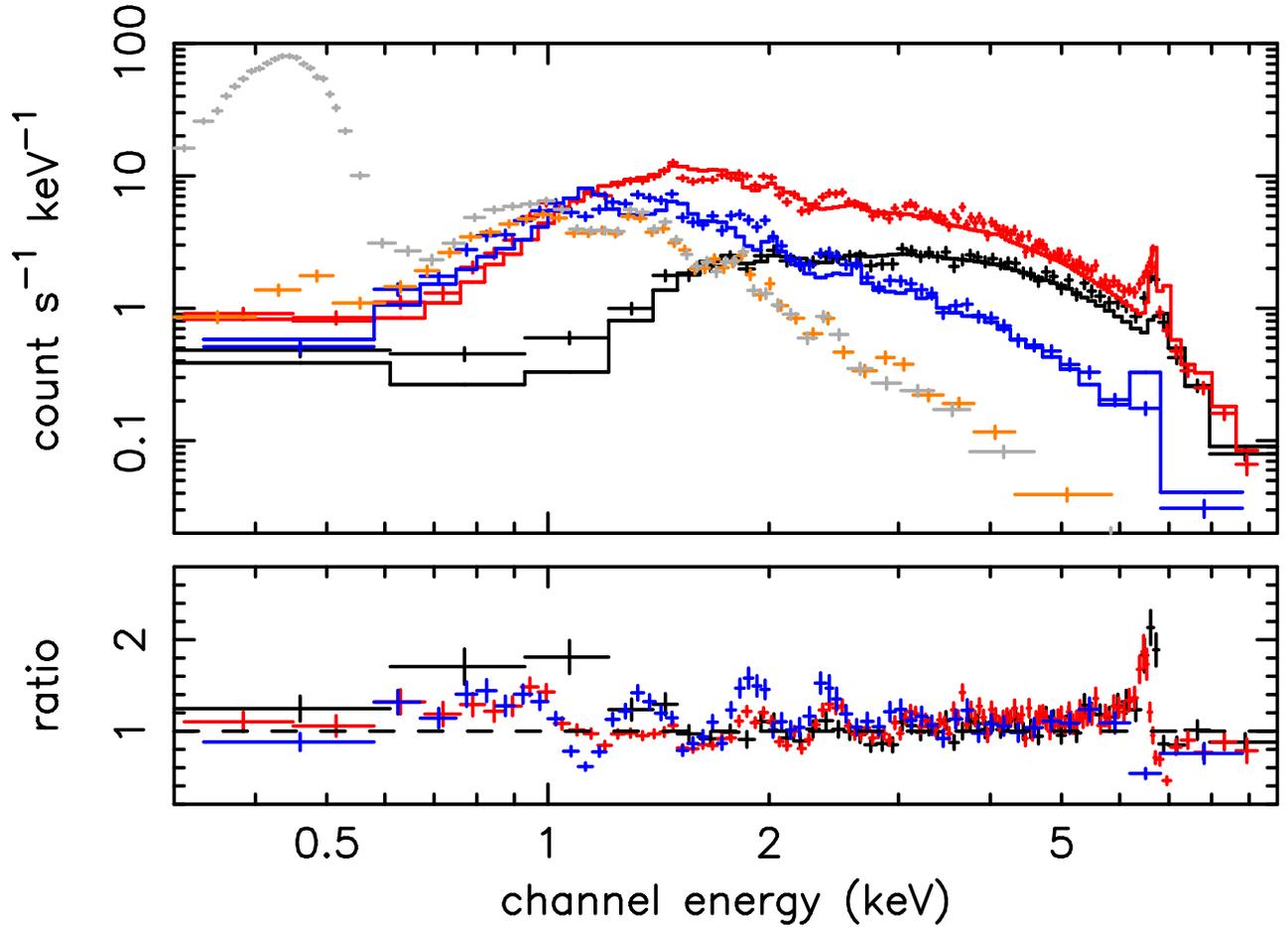}
\caption{Selected {\it Swift} XRT spectra of RS Ophiuchi. Shown are data from $t = $ days 3.17 (black), 5.03 (red), 13.6 (blue), 25.99 (orange), and 29 (grey). The mekal, optically thin, collisionally-ionized plasma emission model fits shown are for days 3.17, 5.03 and 13.6, with their residuals plotted in the bottom panel. \label{spectra}}
\end{figure}

Figure \ref{spectra} shows a sample of the resulting spectra at selected epochs and Table \ref{tab1} gives further information, including the results of the model fits described below. The initial count rate at $t = 3.17$ days doubled in less than two days to peak at around 30 counts s$^{-1}$, then fell again until the emergence of a new soft component beginning at Observation 8 ($t = 26$ days) which dominated the XRT spectrum by Observation 9 \citep[$t = 29$ days]{osb06a} - see Figure \ref{spectra}. The gross evolution of the spectrum was a progressive softening during the period of these X-ray observations. 






\section{Phases of Remnant Evolution}

In the earliest phase of the outburst, the ejecta from the WD surface are expected to traverse the binary system on a timescale $t \sim 2$d for $v_{ej} = 3000$ km s$^{-1}$ \citep{dob94,fek00}. Subsequent to this, using models previously applied to the evolution of SNR, \citet{bod85} found that Phase I (where the ejecta were still important in supplying energy to the shocked stellar wind) ended in the first few days after the 1985 outburst. Indeed, using the parameters derived by \citet{obr92} in Equations 5 and 9 of \citet{bod85}, we derive $t \simeq 6.2$d for the duration of Phase I (see Figures \ref{NH} and \ref{vs}). Thereafter, the remnant evolution can be well modelled by the instantaneous release of energy at a point. The \citet{bod85} analytical model of the 1985 outburst led to the conclusion that at the time the EXOSAT observations began ($t = 55$d), the remnant was in the transition between Phase II (where a blast wave is being driven into the wind and is so hot as to be effectively adiabatic) and Phase III (where the shocked material is well cooled by radiation). Initially, a double shock system is established, with the forward shock being driven into the stellar wind and a reverse shock being driven into the otherwise unshocked ejecta.  In Phase II, the Primakoff similarity solution for this type of explosion into an $r^{-2}$ density distribution (where $r$ is the radial distance from the site of the explosion) gives

\begin{equation}
r_s = a t^{2/3}
\end{equation}

where $r_s$ is the shock radius and $a$ is a parameter which is a function of the energy $E$ communicated by the ejecta to the shocked wind, the mass loss rate in the red giant wind $\dot M$ and the red giant wind velocity $u$. Similarly, in Phase III, 

\begin{equation}
r_s = b t^{1/2} .
\end{equation}

For a strong shock, the post-shock temperature $T_s$ is given by 

\begin{equation}
T_s = \frac{3}{16} \frac{\bar m v_{s}^{2}}{k}
\end{equation}

where $k$ is Boltzmann's constant and $\bar m = 10^{-24}$ g is the mean particle mass, including electrons.

The \citet{obr92} analytical models found good fits to the evolution of the remnant up to around $t = 70$ days after outburst when there was a significant decline in soft X-ray flux. This was explored in terms of numerical models of the forward shock running off the end of the red giant wind re-established between the 1967 and 1985 outbursts, as suggested by \citet{mas87}. However, \citet{obr92} noted that discrepancies between the observed X-ray and models, particularly evident at low energies, could be explained by a contribution from unveiling of the central nuclear burning source (see below). The final EXOSAT observation on day 251 was proposed as being consistent with low luminosity remnant nuclear burning on the WD surface \citep{mas87}.

\section{Interpreting the Observations of the 2006 Outburst}

In the 2006 outburst, the {\it Swift} XRT observations reported here started towards the end of the period in which \citet{bod85} concluded that the reverse shock running into the ejecta may still be important (Phase I), then cover the period of the expected Phase I/II transition and extend into that of Phase II/III. 


From Table \ref{tab1} and Figure \ref{spectra} it can be seen that until around $t = 26$ days (Observation 8) the observed X-ray flux at first rose rapidly, then gradually declined, becoming progressively softer with time. However, as noted above, an excess at energies below around 0.7 keV was evident from $t \sim 26$d onwards and this varied dramatically thereafter \citep{osb06b}. A fuller exploration of the evolution of the soft component will be given in \citet{osb06}.

Extensive model fitting of the early XRT results has been undertaken. The presence of obvious line emission, both in the XRT spectrum and from a higher resolution {\it Chandra} X-ray observation on 2006 February 26 \citep{nes06}, strongly suggested that models of emission from a high temperature thermal plasma were most appropriate. Our first order approach has therefore been to fit a single temperature mekal model to the emission using XSPEC, yielding both line and continuum radiation, with the thermal bremsstrahlung contribution being increasingly dominant at higher temperatures. 

Up to and including Observation 7 at $t = 18.17$ days, the fit was performed over the whole 0.3 - 10 keV range of the XRT data. After this date, only data for $E > $0.7 keV were used in order to separate out emission from the additional, highly variable, soft component seen after this time. We have constrained the interstellar column to be $[N_H]_{IS} = (2.4 \pm 0.6) \times 10^{21}$ cm$^{-2}$ throughout \citep{hje86} but we let the additional column from the overlying red giant wind, $[N_H]_W$, be a free parameter up to and including Observation 7. $[N_H]_W$ was then fixed at $1.8 \times 10^{21}$ cm$^{-2}$ thereafter as the new soft X-ray emission component began to dominate the emission at low energies. Spectral fits at several epochs using a solar abundance model are shown in Figure \ref{spectra}. We note that elemental abundance enhancements were required by \citet{bod85} in their model and also by \citet{sni87} and \citep{sho96} from IUE observations of the 1985 outburst. However, we choose solar abundances until we have performed detailed analyses of the evolution of the shell (as described further in  the Discussion section below). We also note that these single temperature fits, as may be reasonably expected, fail to reproduce all of the spectral detail and should not be overly interpreted at this stage.

In the first three epochs, (Observations 1, 2a and 2b) the Fe K line was clearly detected at around 6.7 keV, as expected for the fitted plasma temperatures. In the first two epochs, there was also obvious excess emission at lower energies in the line compared to the single temperature mekal model fit. This could be characterised by an additional line at 6.4 keV which suggests X-ray reflection from a cold (or moderately ionised) gas as seen in some AGN and magnetic CVs \citep{bea95,don95}. We consider scattering in the red giant wind more 
likely in this case, a situation well documented in a number of supergiant X-ray binaries (e.g. \citet{van05} and references therein).
The 
equivalent width of this line ($ \sim 70$ eV in Observation 2a) is consistent 
with fluorescent emission by the surrounding gas with a column density 
$\sim 6\times 10^{22}$ cm$^{-2}$ \citep{mak85}.

\begin{figure}
\centering
\includegraphics[width=0.8\textwidth,angle=-90]{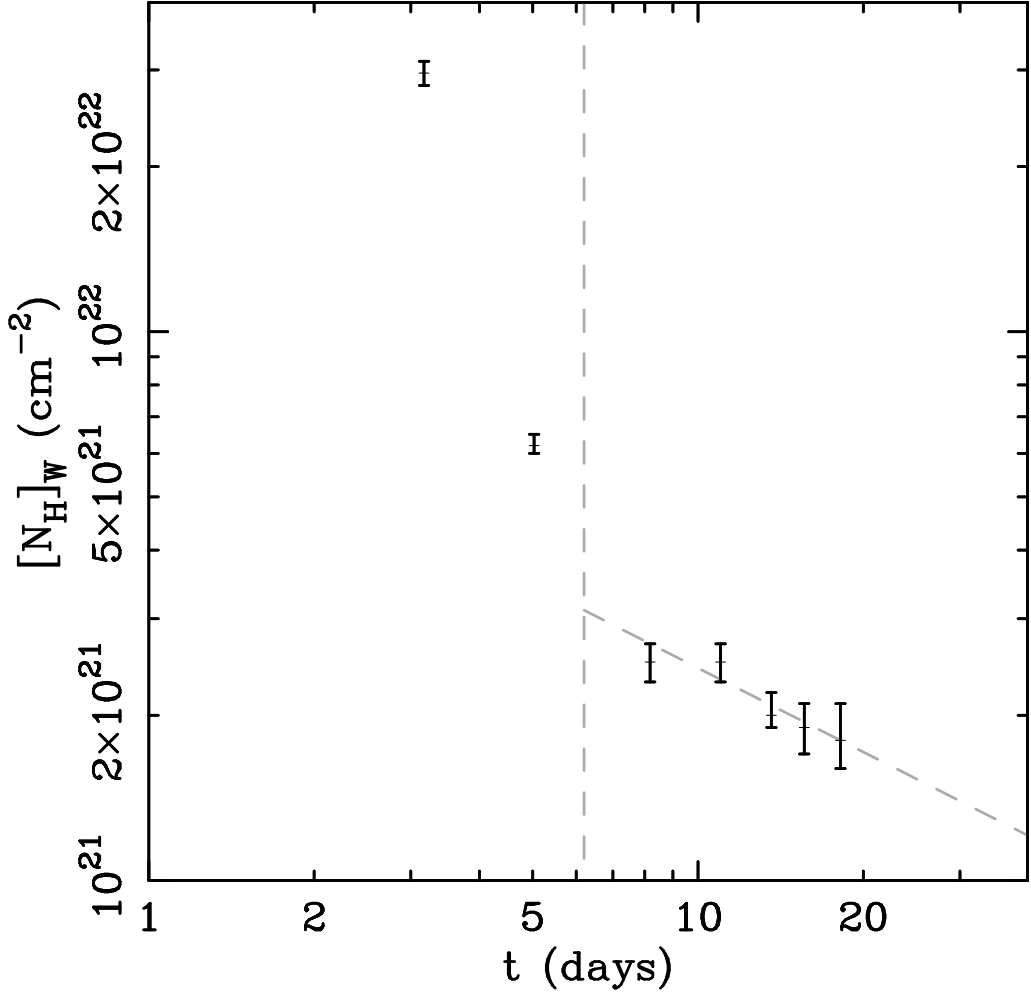}
\caption{Temporal behaviour of the overlying absorbing column in the red giant wind, $[N_H]_W$, net of the assumed interstellar column of $2.4 \times 10^{21}$ cm$^{-2}$, derived from the XRT spectral fits. A vertical dashed line on this Figure (and Figure \ref{vs}) indicates the expected end of Phase I development. The diagonal dashed line is a power-law of index $\alpha = 0.5$ . Note that Observation 8 is not shown as $[N_H]_W$ was fixed at the value derived for Observation 7 (see text for details). \label{NH}}
\end{figure}

\begin{figure}
\centering
\includegraphics[width=0.9\textwidth,angle=-90]{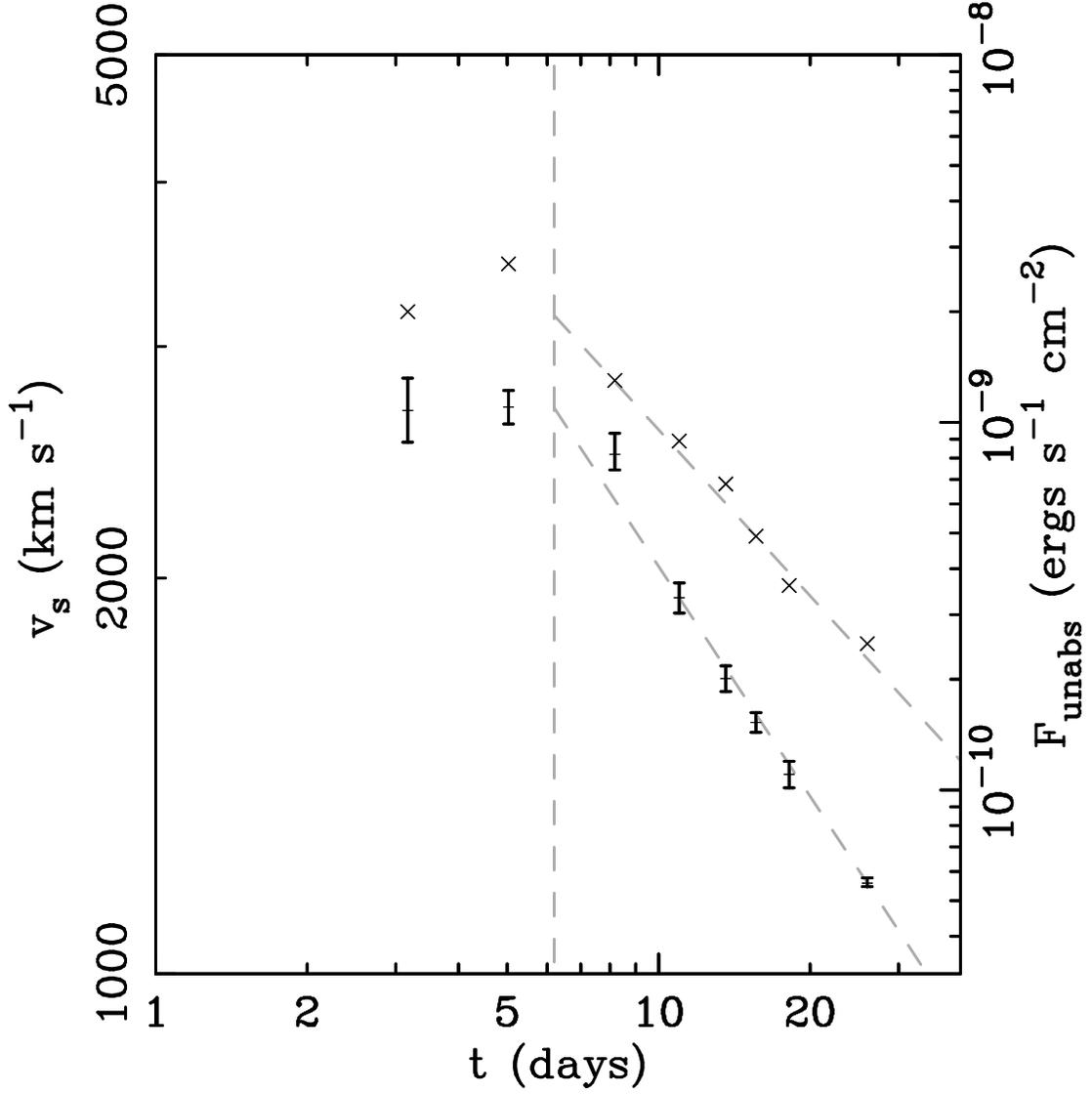}
\caption{The XRT-derived temporal behaviour of the shock velocity, $v_s$ (points with error bars) and unabsorbed (0.7 - 10 keV) flux, $F_{unabs}$ (crosses). The diagonal dashed lines are power-laws with index $\alpha = 0.6$ (for $v_s$ points) and $\alpha = 1.5$ (for $F_{unabs}$ - see text for details). \label{vs}}
\end{figure}

Figures \ref{NH} and \ref{vs} illustrate the results for important parameters derived from the fits. For example, Figure \ref{NH} shows a monotonic decline in $[N_H]_W$ with time, in a manner qualitatively expected as the forward shock traverses the overlying wind. The forward shock velocity, $v_s$, has been derived from the best fit plasma temperature and Equation 3 above. It can also immediately be seen that the implied shock velocities are of the same order as those derived spectroscopically \citep{bui06,eva06,nes06}. The unabsorbed X-ray flux, $F_{unabs}$, shows an initial rise to a peak at around $t = 5$d (with $L_X \approx 1.2 \times 10^{36}$ erg s$^{-1}$  between 0.3 and 10 keV at $d = 1.6$ kpc) followed again by a subsequent decline.

We can also undertake a more quantitative exploration of the basic model from these results in combination with VLBA observations obtained on 2006 February 26 \citep{obr06,obr06a}. These were taken almost simultaneously with Observation 5 of {\it Swift} (day 13.6), and showed a clumpy ring of emission which, if associated with the forward shock, would give a radius of approximately $2.1 \times 10^{14}$ cm at $d = 1.6$ kpc. The value of $N_H$ in the overlying wind (the net neutral column derived from the fits and shown in Table \ref{tab1} and Figure \ref{NH}) can be found from

\begin{equation}
[N_H]_W = \frac{X}{4 \pi m_H} \frac{\dot M}{u} [\frac{1}{r_s} - \frac{1}{r_{out}}]
\end{equation}

where $X$ is the mass fraction of hydrogen in the wind, $m_H$ is the mass of a hydrogen atom, $u$ is the wind speed and $r_{out}$ is the outer wind radius ($= u \Delta t$, where $\Delta t$ is the time since the last outburst, $\sim 21.04$ years). With $X = 0.7$ for solar composition, $u = 20$ km s$^{-1}$ and $\dot M/u$ from \citet{obr92}, this gives $[N_H]_W = 9\times10^{20}$ cm$^{-2}$, i.e. within a factor of two of the derived value from the single temperature model fits at this time (see Table \ref{tab1}). We note that $[N_H]_W$ derived in Equation 4 will be an underestimate for the value of the column for the whole shocked shell as it is calculated from the line-of-sight (front) edge. In addition, X-ray estimates of $[N_H]_W$ depend on the abundances in the wind (we have assumed these are solar) and furthermore \citet{bau06} suggest that $[N_H]_{IS}$ derived from X-ray observations may differ somewhat from 21 cm values as it also samples H$_2$ in the ISM.

After $t \sim 6$ days, (by which time the \citet{bod85} model expected Phase I to be effectively over) it can be seen from Figure \ref{vs} that $v_s$ declines monotonically with time, following a power law decay from $t \sim 10$ days onwards. Here, $v_s \propto t^{-\alpha}$, where $\alpha \simeq 0.6$. This compares more precisely to the expected value of $0.5$ for Phase III, rather than  $\alpha = 0.33$, expected in Phase II evolution \citep{bod85}. Equation 4 implies that $[N_H]_W \propto {t}^{-\alpha}$ at early times (as $r_{out} = 1.3\times10^{15}$ cm $\gg r_s$), where $\alpha = 0.67$ (Phase II) or $0.5$ (Phase III) respectively. From Figure \ref{NH}, it appears that after the implied end of Phase I, the data can in fact be better fitted by $\alpha = 0.5$. Finally, after the expected end of Phase I, $F_{unabs}$ decayed as $t^{-1.5}$, as expected from a well-cooled shock.

The \citet{sok06} observations of RS Oph with the RXTE PCA between days 3 and 21 are also consistent with the basic shock model. Their results imply Phase II evolution after the first few days from the behaviour of their derived temperature, but are more consistent with Phase III evolution during this period from the behaviour of the X-ray flux. The usable energy range of the PCA (2-25 keV) means however that their observations were less sensitive to $N_H$ and they did not detect the emergence of the soft component.

At $t = 13.6$ days, our fits to the {\it Swift} data imply $v_s \simeq 1700$ km s$^{-1}$. We note that we could obtain complete consistency between the VLBA imagery and simple derivation of $v_s$ from the {\it Swift} model fits if $a = 2.7 \times 10^{10}$ cm s$^{-2/3}$ and a larger distance of $d = 2.4$ kpc, if the remnant were behaving precisely as one would expect for the Primakoff (Phase II) solution throughout. However, the time dependence of $v_s$, $[N_H]_W$ and $F_{unabs}$ are more like what one would expect in Phase III, but in this case $b = 3.7 \times 10^{11}$ cm s$^{-1/2}$ and for consistency with the VLBA results, the distance would have to be increased to 3 kpc. Such high values of the distance were accepted prior to 1985 \citep{bods87}. On the other hand, \citet{sni87} notes that the failure to detect material at the typical velocities of the Carina Arm in {\it IUE} spectra places an upper limit on the distance to RS Oph of about 2 kpc. In addition of course, we have shown that there is a transition at around $t = 6$ days between remnant phases. Thus although the simple analysis is supportive of the qualitative correctness of the current models, as outlined below, detailed physical modelling is required to build a fully self-consistent picture.



\section{Conclusions and future work}

Our X-ray observations of the very earliest phases of the 2006 outburst of RS Oph are broadly consistent with the basic model of remnant evolution proposed by \citet{bod85} and further explored by \citet{obr92}. In particular, it appears that Phase I may have ended by $t \sim 6$d as predicted. However, our first-order analysis suggests that the remnant moved rapidly into Phase III in this outburst.

Fits to the XRT data have also been attempted using vmekal where the abundances have been allowed to vary. For the single temperature fit, this resulted in a trend of decreasing elemental abundance enhancement with time, as would be expected as emission from the enriched nova ejecta at early times becomes increasingly less significant than that from the generally less enriched red giant wind. A series of fits using multi-temperature mekals was also performed producing improved fits to the data. However, detailed models involving full hydrodynamical simulations are ultimately required to fit the X-ray data in a physically meaningful way. These simulations produce an evolving range of temperatures and densities with radius from the central source (see \citet{obr92}). Resonant scattering 
of X-rays in the red giant wind may also be occurring, and therefore some of the 
X-ray emission line flux may be due to photoionization of this gas by 
the X-ray emitting shock. In addition, VLBI results \citep{obr06a} have shown that assumptions of spherical symmetry are called into question, and higher spatial dimension  modelling may be required. We plan this more detailed modelling to be the subject of a future paper.

In future papers we will report the results of our continued monitoring of the source with {\it Swift}. As well as attempting to understand the origin of the highly variable soft X-ray component (to be discussed in \citet{osb06}, and which may possibly arise from continued nuclear burning near the Eddington luminosity on the WD surface), we are particularly interested in detailed investigation of three subsequent predicted phases of development: (a) the epoch when the forward shock is forecast to reach the end of the red giant wind at around $t = 80$ days, (b) the timescale for turn-off of the high luminosity phase of the TNR, and (c) the re-establishment of accretion in the central binary and the overlying wind from the giant.  


\acknowledgments

The authors are very grateful to the {\it Swift} Mission Operations Center staff for their superb efforts in supporting the observations reported here. We also acknowledge quick-look results provided by the ASM/RXTE team. P. Caraveo and A. de Luca contributed to work on the BAT data. SS gratefully acknowledges partial support from NSF and NASA grants to ASU. SPSE acknowledges the support of the UK's Nuffield Foundation. JPO, KLA, APB and MRG acknowledge support from PPARC. J-UN is supported by the {\it Chandra} Fellowship Program. We also wish to thank an anonymous referee for their very helpful comments on the original version of the manuscript.



{\it Facilities:} \facility{Swift(BAT,XRT)} \facility{RXTE (ASM)}




\clearpage
\thispagestyle{empty}

\begin{deluxetable}{lcccccccc}
\rotate
\tablecaption{Summary of RS Oph XRT Observations and Spectral Fits. \label{tab1}}
\tablewidth{0pt}
\tablehead{
\colhead{Obs.} &
\colhead{Date (Day)} &
\colhead{Exposure} &
\colhead{Count rate} &
\colhead{kT\tablenotemark{a}} &
\colhead{Correction} &
\colhead{[N$_{H}$]$_W$\tablenotemark{b}} &
\colhead{$\chi^{2}$/dof} &
\colhead{Unabs. Flux\tablenotemark{c}}\\
\colhead{} &
\colhead{} &
\colhead{(s)} &
\colhead{(count s$^{-1}$)} &
\colhead{(keV)} &
\colhead{Factor\tablenotemark{a}} &
\colhead{(10$^{22}$ cm$^{-2}$)} &
\colhead{} &
\colhead{(erg cm$^{-2}$ s$^{-1}$)}
}
\startdata
001 & 2006-02-16T00:05 (3.17)& 664 & 14.2~$\pm$~0.2 & 8.44$^{+0.98}_{-0.91}$  &
 1.2 & 2.96~$\pm$~0.15 & 460/286 &
 2.0~$\times$~10$^{-9}$ \\
 002a & 2006-02-17T20:33 (5.03) & 999 & 31.5~$\pm$~0.2 & 8.54$^{+0.50}_{-0.49}$ &
1.1  & 0.62$^{+0.03}_{-0.02}$  & 993/585 &
 2.7~$\times$~10$^{-9}$\\
 002b & 2006-02-21T00:10 (8.18)& 844 & 19.8~$\pm$~0.2 & 7.24$^{+0.54}_{-0.39}$ &
1.0 &  0.25~$\pm$~0.02 & 631/436 &
 1.3~$\times$~10$^{-9}$\\
 004 & 2006-02-23T19:37 (10.99)& 950 & 15.5~$\pm$~0.2 & 4.38~$\pm$~0.23 &
1.3 & 0.25~$\pm$~0.02 & 484/327 &
 8.9~$\times$~10$^{-10}$ \\
 005 & 2006-02-26T10:18 (13.6)& 897 & 12.3~$\pm$~0.1 & 3.30~$\pm$~0.15 &
1.1  & 0.20$^{+0.02}_{-0.01}$ & 465/289 &
 6.8~$\times$~10$^{-10}$\\
 006 & 2006-02-28T10:31 (15.61)& 1041 & 10.6~$\pm$~0.1 & 2.83$^{+0.10}_{-0.09}$ &
1.0 & 0.19~$\pm$~0.02 & 559/283 &
 4.9~$\times$~10$^{-10}$\\
 007 & 2006-03-03T00:04 (18.17)& 738 & 8.6~$\pm$~0.1 & 2.36$^{+0.11}_{-0.12}$ &
 1.0 & 0.18$^{+0.03}_{-0.02}$ & 353/206
& 3.6~$\times$~10$^{-10}$\\
 008\tablenotemark{d} & 2006-03-10T19:38 (25.99) & 1605& 6.5~$\pm$~0.1 & 1.61$^{+0.03}_{-0.02}$ & 1.1 & 0.18 & 708/189 &  2.5$\times$10$^{-10}$\\
\enddata
\tablenotetext{a}{Factor by which the count rate (and flux) has been corrected to account for bad columns on the CCD. This varies due to the changing placement of the source on the detector.}
\tablenotetext{b}{Absorption above the ISM value of 2.4~$\times$~10$^{21}$ cm$^{-2}$.}
\tablenotetext{c}{0.7-10 keV. Bolometric corrections range from 1.5 to 1.7 for Observations 1 to 8.}
\tablenotetext{d}{Fits to 0.7 - 10 keV and additional column fixed at previous value (see text).}
\end{deluxetable}

\clearpage

\end{document}